\documentclass[aps, prd, twocolumn,print, showpacs, nofootinbib]{revtex4}
\usepackage{txfonts}
\usepackage{amssymb}
\usepackage{mathrsfs}
\usepackage{graphicx}
\usepackage{epsfig}
\usepackage{dcolumn}
\usepackage{bm}

\begin{document}

\title{Could strange stars be in the color-flavor-locked phase: Tested by their thermal evolutions}
\author{Quan Cheng}{}
\author{Yun-Wei Yu}\email{yuyw@phy.ccnu.edu.cn}
\author{Xiao-Ping Zheng}{}

\affiliation{Institute of Astrophysics, Central China Normal
University, Wuhan 430079, China}

\date{Dec 2012}
\begin{abstract}
The thermal evolution of strange stars in both normal and
color-flavor-locked (CFL) phases are investigated together with the
evolutions of the stellar rotation and the r-mode instability. The
heating effects due to the deconfinement transition of the stellar
crust and the dissipation of the r-modes are considered. As a
result, the cooling of the stars in the normal phase is found to be
not very different from the standard one. In contrast, for the stars
in the CFL phase, a big bump during the first hundred years and a
steep decay ($\sim$7\% in ten years) at the ages of $\sim10^{4-6}$
yrs are predicted in their thermal evolution curves. These unique
features provide an effective observational test for determining
whether or not the CFL phase is reached in strange stars. This
thermal test method is independent of and complementary to the
rotational test method, which is a direct consequence of the r-mode
instability [see J. Madsen, Phys. Rev. Lett. 85, 10 (2000)].
\end{abstract}

\maketitle

\section{INTRODUCTION} \label{S:intro}

Strange quark matter (SQM), made up of roughly equal numbers of up,
down, and strange quarks, could be the absolute ground state of
strong interaction \cite{Bodmer:1971,Witten:1984,Madsen:1999}, the
most important astrophysical consequence of which is that compact
stars could have a quark matter core (i.e., hybrid star) or even
consist purely of SQM (i.e., strange star, SS;
\cite{ALcock:1986,itoh:1970}). Furthermore, phenomenological and
microscopic studies suggest that SQM at a sufficiently high density
could undergo a phase transition into a color superconducting state,
the typical cases of which are the two-flavor color
superconductivity (2SC) and color-flavor-locked (CFL) phases
\cite{Shovkovy:2005,Alford:2004}. In the following content, we adopt
the abbreviations 2SSs and CSSs for the two types of SSs in the two
different color superconducting phases, respectively, and NSSs for
normal SSs.

Madsen \cite{Madsen:2000} suggested that the SS models can be
constrained by the most rapidly rotating pulsars in low mass X-ray
binaries (LMXBs), because the rotation of compact stars should be
suppressed by r-mode instability, which arises due to the action of
the Coriolis force with positive feedback increasing gravitational
radiation (GR) \cite{Andersson:1998,Frindman:1998}. As a result, it
is found that the CSS model could be not permitted by the LMXB data
because of the too weak viscosities of CSSs, which cannot inhibit
the r-modes effectively. In this paper we suggest something
different from the rotational test to the models---a thermal test
method, i.e, probing the interior of SSs by the temperature
observations of compact stars. The most essential point of our work
is to carefully investigate the influence of the r-mode instability
on the thermal evolution of SSs. Therefore, our independent thermal
test can be complementary to the direct rotational test.

The cooling behaviors of SSs have already been extensively
investigated, ever since the emergence of the SS model. In the
earliest works, it was accepted that the surface temperature of SSs
should be significantly lower than traditional neutron stars at the
same age due to the quark direct Urca (QDU) processes
\cite{Alcock:1988,Pizzochero:1991,Page:1992,Schaab:1996}. However,
by considering that the electron fraction of SQM could be small or
even vanish and thus the QDU processes could be switched off, Schaab
\emph{et al}. \cite{Schaab:1997a,Schaab:1997b} proposed that the
cooling of SSs dominated by the quark-modified Urca and quark
bremsstrahlung processes can be slower than neutron stars with
standard cooling. The possible existence of the color
superconductivity makes the problem more complicated. It has been
shown that the cooling of 2SSs is compatible with existing x-ray
data but that CSSs cool down too rapidly, which completely disagrees
with the observations \cite{Blaschke:2000}.\footnote{In the CFL
case, both the neutrino emission and the quark thermal capacity are
blocked off, whereas the cooling due to the photon emission from the
stellar surface is still effective. Therefore, an extremely large
temporal derivative of temperature can be obtained by dividing the
photon luminosity by the tiny heat capacity, which leads to the
rapid cooling. See the grey band in the right panel of Fig.
\ref{fig:3}.}

Besides the stellar core consisting of SQM, SSs could also sustain a
tiny nuclear crust with a maximum density below neutron drip
($\sim4\times 10^{11}$ g cm$^{-3}$)
\cite{ALcock:1986,Usov:2004,zheng:2006a}. The mass of the crust as a
function of the spin frequency $\nu$ can be expressed as
\cite{Zdunik:2001}
\begin{equation}
M_{\rm c}=M^{0}_{\rm
c}\left(1+0.24\nu^{2}_{3}+0.16\nu^{8}_{3}\right),\label{mc}
\end{equation}
where $\nu_{3}=\nu/10^{3}$ Hz, and $M^{0}_{\rm
c}\approx10^{-5}M_{\odot}$ is the mass of the crust in the static
case. With the spin down of a SS, the star shrinks and the density
at the bottom of the crust exceeds neutron drip. As a result, the
most inner part of the crust falls into the quark core and dissolves
from baryons into quarks. During this process the released binding
energy can heat the star. Therefore, Yuan and Zhang \cite{Yuan:1999}
proposed a so-called deconfinement (DC) heating mechanism, which can
delay the cooling of SSs significantly. Especially for CSSs, the
stellar thermal evolution could be completely determined by the DC
heating \cite{Yu:2006}. As analyzed, the temporal dependence of the
heating power should be determined by the spin-down history.

In \cite{Yuan:1999} and \cite{Yu:2006}, as usual, the spin down of
SSs is considered to be due only to the magnetic dipole radiation
(MDR). However, as aforementioned, the spin of young compact stars
could also be limited by r-mode instabilities. In other words, if
the r-mode amplitude is large enough, the early spin down of SSs
could be dominated by GR rather than MDR \cite{Owen:1998}. This
could significantly influence the process of the release of the
crust-binding energy. On the other hand, as an extra heating effect,
the shear viscous dissipation of r-modes can also deposit energy
into the thermal state of the stars
\cite{Andersson:2001,Watts:2002,Zheng:2006}. The main purpose of
this paper is to test the CSS model by investigating its thermal
evolutions with the above-mentioned heating effects.

In the next section we briefly introduce the thermal evolution
equation of SSs, where both DC heating and the heating due to
viscous dissipation of r-modes (RM heating hereafter) are involved.
The spin evolution of SSs involving both MDR and GR brakings is
described in Sec. \ref{Section III}, which is essential to solve the
thermal evolution equation. The calculated results and some
discussions are provided in Secs. \ref{Section IV} and \ref{Section
V}, respectively.

\section{THERMAL EVOLUTION EQUATION}
For simplicity, in this paper, the temperature in the interior (the
core) of a SS is assumed to distribute nearly uniformly, and the
gradient from the interior temperature to the surface temperature is
considered to be mainly caused by the thin crust. Following
\cite{Gudmundsson:1983}, the relationship between the surface
temperature $T_s$ and the interior temperature $T$ can be expressed
by $T_{s}=3.08\times10^6g_{s,14}^{1/4}T_{9}^{0.55}$, where
$T_9=T/10^9$ K and $g_{s,14}$ is the proper surface gravity of the
star in units of $10^{14}\rm ~cm~s^{-2}$. As a result, the
luminosity of the surface photon emission of the star can be
calculated to $L_{\gamma}=4\pi R^2\sigma T_s^4$, where $R$ is the
stellar radius and ${\sigma}$ is the Stefan-Boltzmann constant. Of
course, as usually considered, neutrino emission could play a much
more important role in the cooling of a young NSS.

Denoting the neutrino luminosity by $L_\nu$ and the powers of the DC
and RM heatings by $H_{\rm DC}$ and $H_{\rm RM}$, the thermal
evolution of a SS can be solved from
\begin{equation}
C\frac{d T}{d t}=-L_{\gamma}-L_{\nu}+H_{\rm DH}+H_{\rm
RM},\label{Tt}
\end{equation}
where $C$ is the thermal capacity of the star. The expressions of
$C$ and $L_{\nu}$ for normal SQM are taken the same as those in
\cite{Yu:2006} and \cite{Zheng:2006}, which were originally
calculated by Iwamoto \cite{Iwamoto:1982}. In the presence of color
superconductivity, the thermal capacity and neutrino emissions
contributed by the paired quarks are significantly suppressed by
some exponential factors, which can be found in
\cite{Blaschke:2000}.

The power of the DC heating is obviously determined by the
transition rate of the nuclear matter at the bottom of the crust.
Equation (\ref{mc}) further indicates that the transition rate
depends on the variation rate of the spin frequency. Therefore, by
denoting the heat release per dissolved neutron by $q_n$, the power
of the DC heating can be estimated by \cite{Yu:2006}
\begin{equation}
H_{\rm DC}=-q_n\frac{1}{m_{b}}\frac{d M_{\rm c}}{d
{\Omega}}{d\Omega\over dt},
\end{equation}
where $m_{b}$ and $\Omega$ are the mass of baryon and spin angular
velocity of the star, respectively. In the conventional MDR model,
the temporal derivation of the angular velocity can be calculated by
${d\Omega/dt}=-{\Omega/\tau_{\rm m}}$, where the magnetic braking
timescale reads $\tau_{\rm m}=1.69\times
10^{9}B_{12}^{-2}(\Omega/\sqrt{\pi G\bar{\rho}})^{-2}~\rm s$,
$B_{12}$ is the magnetic field intensity in units of $10^{12}$ G,
and $\sqrt{\pi G\bar{\rho}}$ represents the order of the Keplerian
rotation limit with $G$ being the gravitational constant and
$\bar{\rho}$ the mean stellar density. However, by considering the
arising of r-mode instability, which can lead to the loss of stellar
angular momentum via GR, the spin-down history of SSs needs to be
reinvestigated more carefully.

It is natural to consider that a great amount of the stellar
rotational energy can be transferred into r-modes. Subsequently, the
energy carried by the r-modes can be further dissipated by both bulk
and shear viscosities. To be specific, the part of the r-mode energy
dissipated by the bulk viscosity escapes from the star via neutrino
emission, whereas the part corresponding to the shear viscosity can
be deposited into internal energy, which could be of great
importance for the stellar thermal evolution. Following
\cite{Owen:1998} and \cite{Zheng:2006}, the power of such RM heating
can be written as
\begin{equation}
H_{\rm RM}={2\tilde{E}\over \tau_{\rm
sv}}=\frac{\alpha^{2}\tilde{J}MR^{2}\Omega^{2}}{\tau_{\rm sv}},
\end{equation}
where $\tilde{E}={1\over2}\alpha^{2}\tilde{J}MR^{2}\Omega^{2}$ is
the canonical energy of the r-modes with
$\tilde{J}=1.635\times10^{-2}$ for $n=1$ polytrope, $\alpha$ the
amplitude of the r-modes, and $M$ the stellar mass. For normal SQM,
the shear viscous time scale is $\tau_{\rm
sv}=5.41\times10^{9}\alpha_{\rm
c,0.1}^{5/3}T_{9}^{5/3}\hspace{0.1cm}\textrm{s}$, where $\alpha_{\rm
c,0.1}=\alpha_c/0.1$ is the strong coupling
\cite{Lindblom:1999,Madsen:2000}. In presence of color
superconductivity, a fraction or all of the quarks are paired and
thus the shear viscosity contributed by these paired quarks should
be suppressed by a factor of $\exp(\Delta/3k_{B}T)$, where $\Delta$
is the pairing gap. In such a case, the shear viscosity due to
electron-electron scattering becomes more important for the r-mode
damping. The corresponding time scale can be expressed to $\tau_{\rm
sv}^{\rm ee}=2.95\times 10^{7}(\mu_{\rm e}/\mu_{\rm
q})^{-14/3}T_{9}^{5/3}\hspace{0.1cm}\rm s$, where $\mu_e$ and
$\mu_q$ are the chemical potentials of electrons and quarks,
respectively. On the other hand, as pointed out by
\cite{Madsen:2000}, an extra viscosity due to the rubbing between
the solid nuclear crust and the electron atmosphere of the quark
core could lead to a more effective dissipation of the r-mode
energy, which determines a time scale of $\tau_{\rm sr}=1.42\times
10^{8}(\nu/1{\rm ~kHz})^{-1/2}T_{9}\hspace{0.1cm}\rm s$.

\begin{figure*}
\centering
\resizebox{\hsize}{!}{\includegraphics{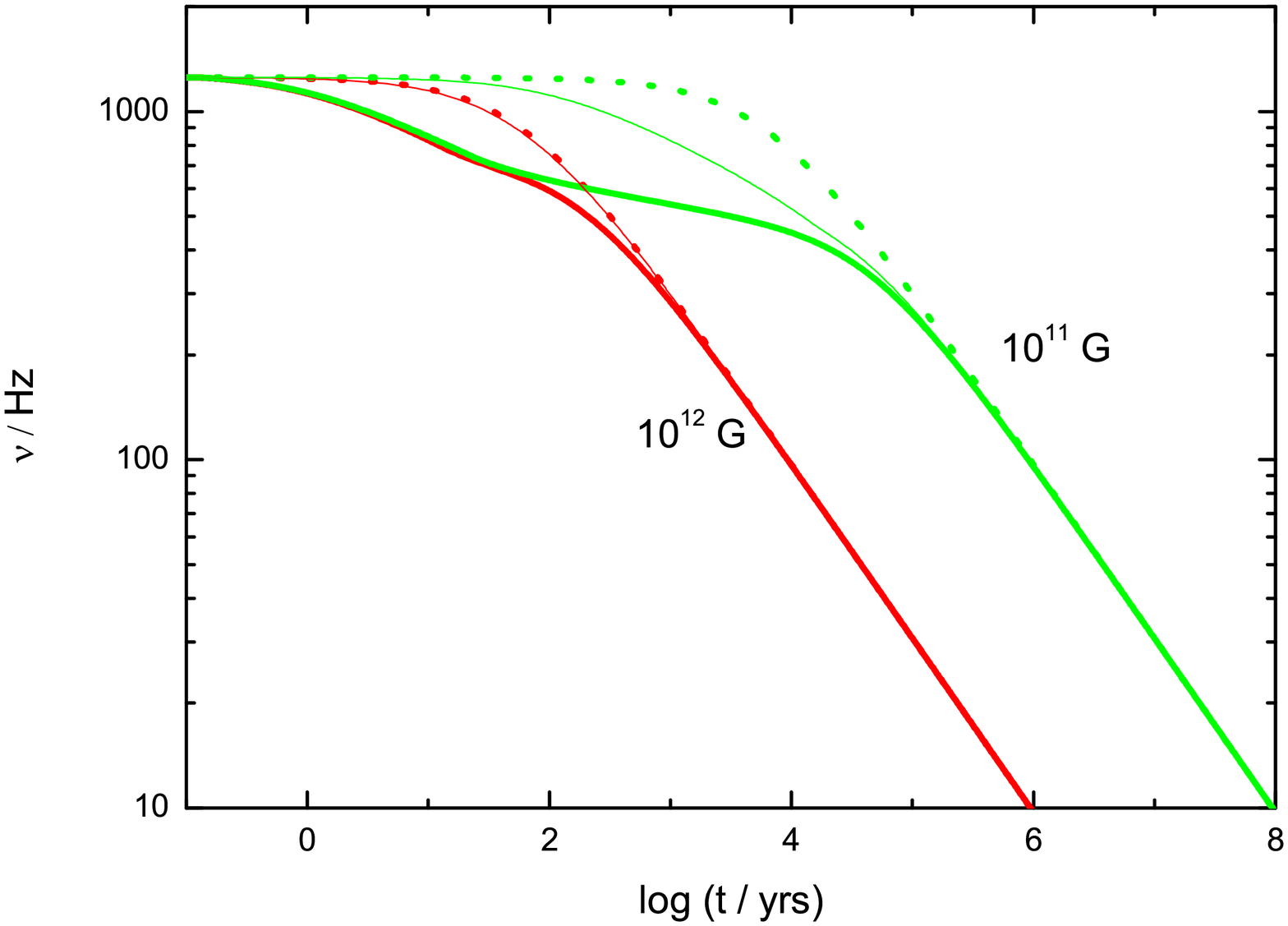}\includegraphics{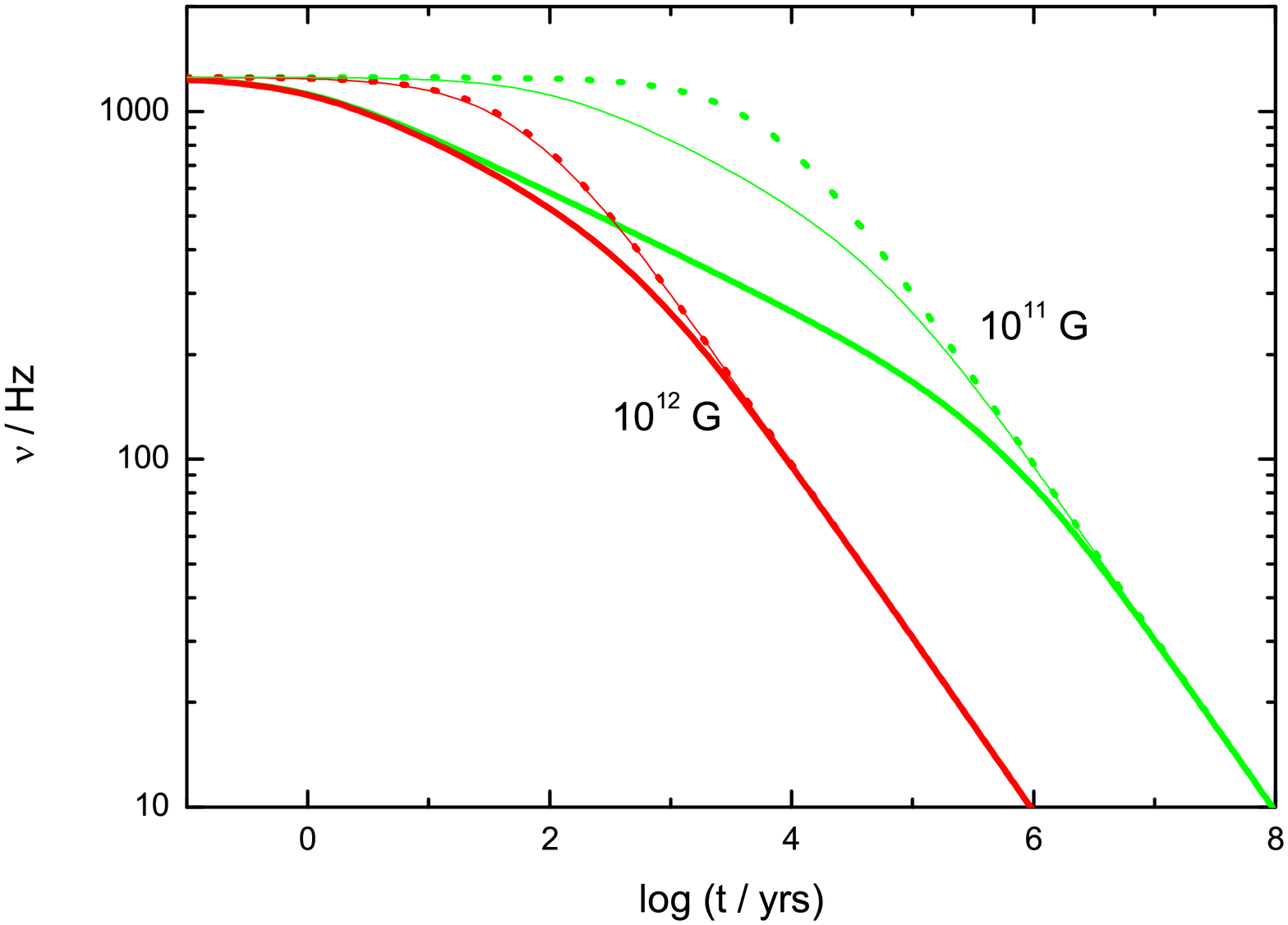}}
         \caption{Spin evolutions of NSSs (left panel) and CSSs (right
         panel). The solid and dotted lines correspond to the case
of GR+MDR braking and of only MDR braking, respectively. Two
different saturation values of $\kappa=10^{-6}$ and $10^{-8}$ are
respectively represented by the thick and thin lines. The typical
magnetic fields of canonical pulsars of $10^{11}$ G and $10^{12}$ G
are taken as labeled.
          }
   \label{fig:1}
\end{figure*}
\section{SPIN DOWN OF SSS} \label{Section III}
As mentioned above, the GR due to r-mode instability could play an
important role in the spin down of a SS. Following the
phenomenological model developed by Owen \emph{et al}.
\cite{Owen:1998} and Ho and Lai \cite{Ho:2000}, the spin down of the
star affected by both MDR and GR brakings can be solved from
\begin{equation}
\frac{d\Omega}{dt}=-\frac{\Omega}{\tau_{\rm
m}}-\frac{2Q\alpha^{2}\Omega}{\tau_{\rm v}},\label{omegat}
\end{equation}
where $Q=0.094$ is a constant determined by the stellar structure.
The viscous time scale here, $\tau_{\rm v}$, includes the effects of
both bulk and shear viscosities, which can be calculated by
$\tau_{\rm v}=\left(\tau_{\rm bv}^{-1}+\tau_{\rm
sv}^{-1}\right)^{-1}$. For normal SQM, the bulk viscous time scale
reads $\tau_{\rm bv}=0.886(\Omega/\sqrt{\pi
G\bar{\rho}})^{-2}T_{9}^{-2}m_{s,100}^{-4}\hspace{0.1cm}\textrm{s}$,
 where $m_{s,100}$ is the mass of
strange quark in units of 100 MeV \cite{Lindblom:1999,Madsen:2000}.
In presence of color superconductivity, this time scale should also
be prolonged by a factor of $\exp(2\Delta/k_{B}T)$. In order to
solve Eq. (\ref{omegat}), we must simultaneously determine the
evolution of the r-mode amplitude by
\begin{equation}
\frac{d\alpha}{dt}=\alpha\left(\frac{1}{\tau_{\rm
g}}-\frac{1-\alpha^{2}Q}{\tau_{\rm v}}+\frac{1}{2\tau_{\rm
m}}\right),\label{alphat}
\end{equation}
where the GR time scale reads $\tau_{\rm g}=3.26(\Omega/\sqrt{\pi
G\bar{\rho}})^{-6}\hspace{0.1cm}\textrm{s}$ for $n=1$ polytrope.

It is generally believed that r-mode instability would finally
arrive at a saturation state due to some nonlinear effects. Lindblom
\emph{et al}. \cite{Lindblom:2001,Lindblom:2002} suggested that the
nonlinear saturation may be determined by dissipation of energy in
the production of shock waves. Moreover, the decay of the amplitude
of the order unity is due to leaking of energy into other fluid
modes, leading to a differential rotation configuration
\cite{Gressman:2002}. Afterwards, the coupling between r-modes and
other modes was analyzed \cite{Arras:2003,Bondarescu:2007}. On the
other hand, the role of differential rotation in the evolution of
r-modes was also studied thoroughly \cite{Sa:2005, Yu:2009}. All
these works obtained a credible conclusion that the maximal
saturation amplitude of r-modes may not be larger than the small
value of $10^{-3}$.

During the saturation state, Eqs.
(\ref{omegat}) and (\ref{alphat}) would be replaced by
\begin{equation}
\frac{d\Omega}{dt}=-\frac{\Omega}{\tau_{\rm m}}\frac{1}{1-\kappa
Q}-\frac{2\Omega}{\tau_{\rm g}}\frac{\kappa Q}{1-\kappa
Q},\label{omegaT}
\end{equation}
and
\begin{equation}
\frac{d\alpha}{dt}=0,
\end{equation}
respectively, where $\kappa=\alpha_{\rm sat}^2$. As a rough
estimation, the upper limit of the r-mode energy can be connected to
the stellar rotational energy by $\tilde{E}_{\max}\simeq \kappa
QE_{\rm rot}$.

\begin{figure*}
\centering
\resizebox{\hsize}{!}{\includegraphics{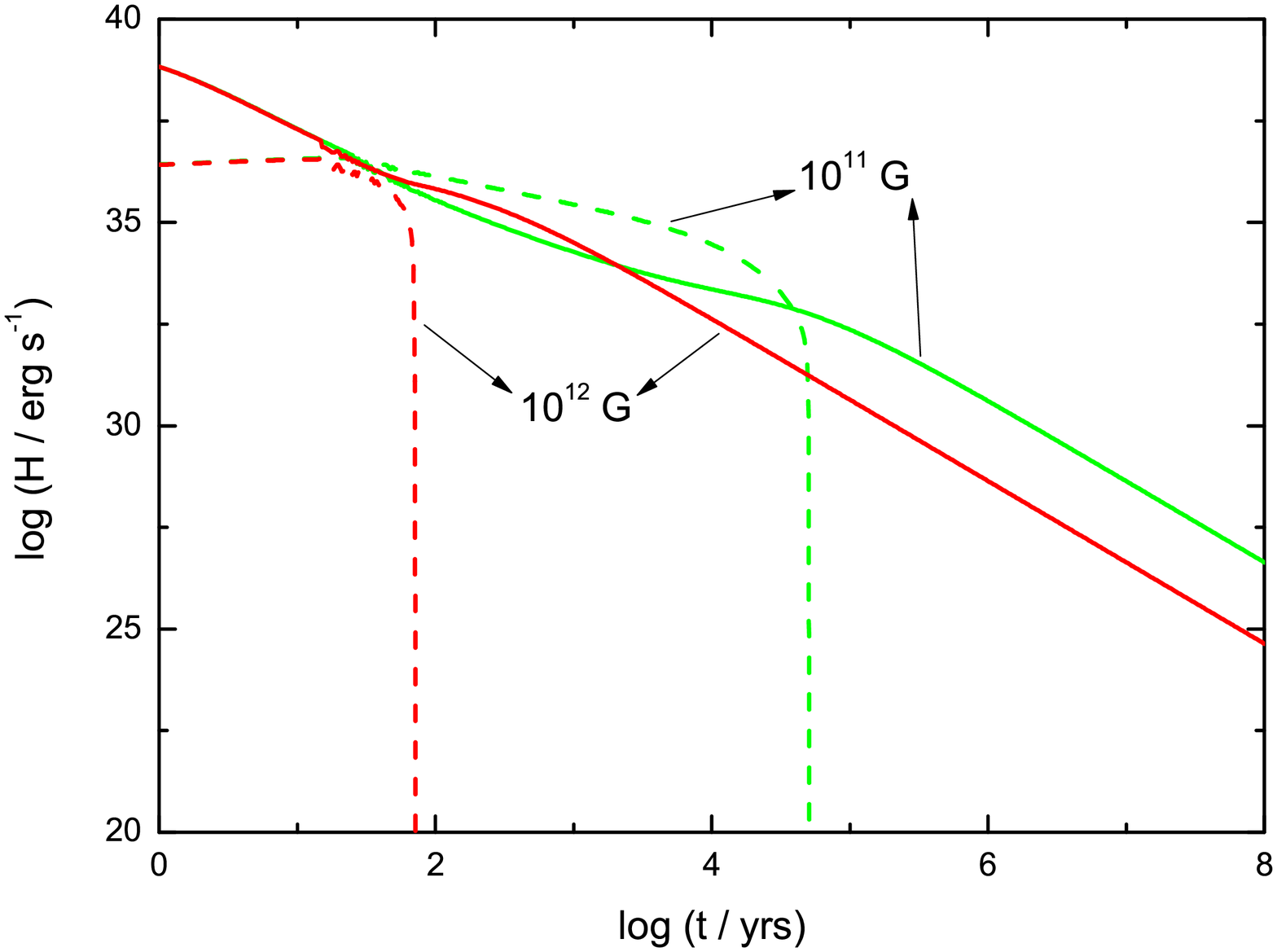}\includegraphics{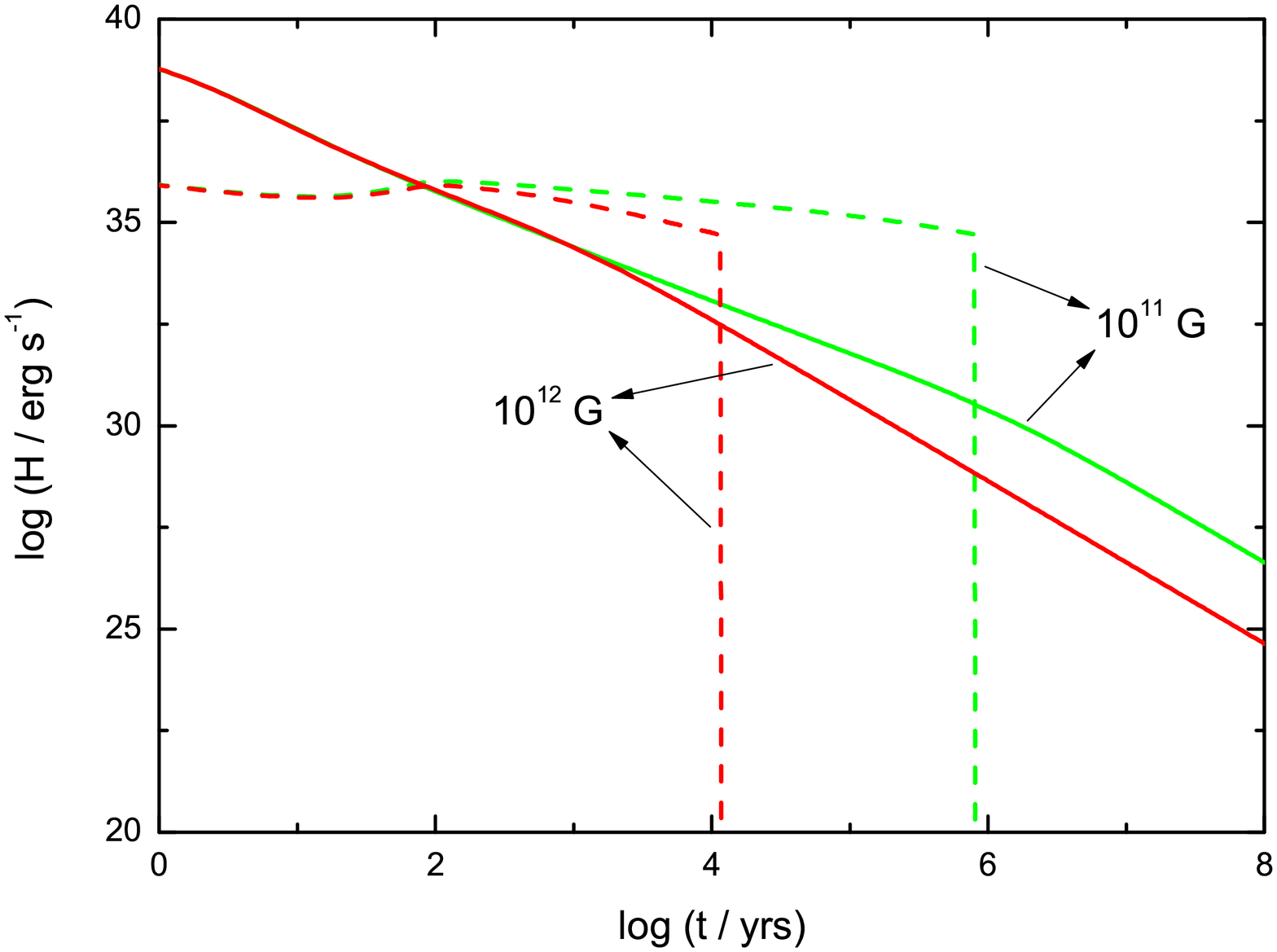}}
         \caption{The temporal dependences of the powers of the DC (solid) and
         RM (dashed) heatings for NSSs (left panel) and CSSs (right panel). The magnetic fields
         are the same as in Fig. \ref{fig:1}.
         }
   \label{fig:2}
\end{figure*}
\begin{figure*}
\centering
\resizebox{\hsize}{!}{\includegraphics{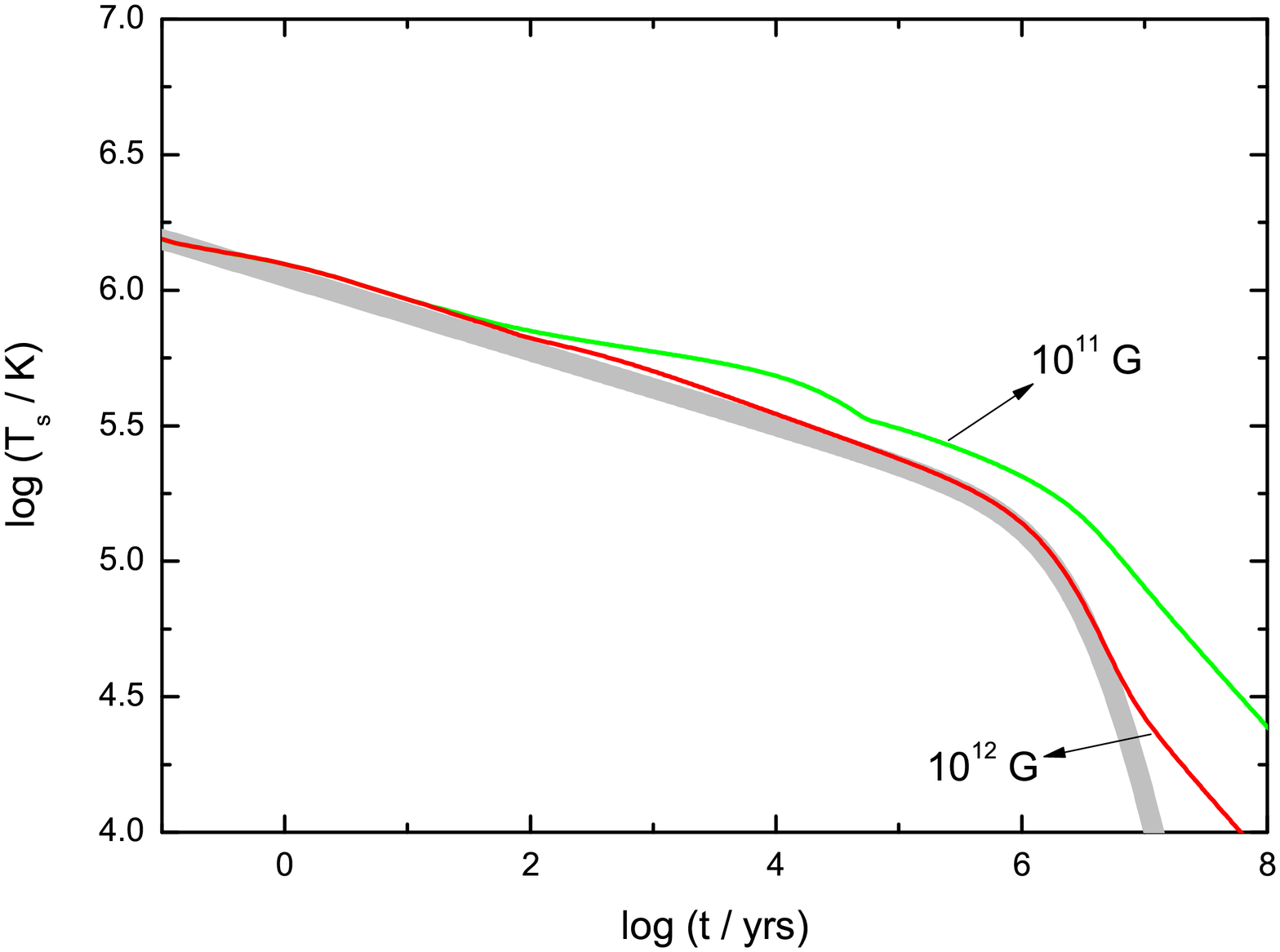}\includegraphics{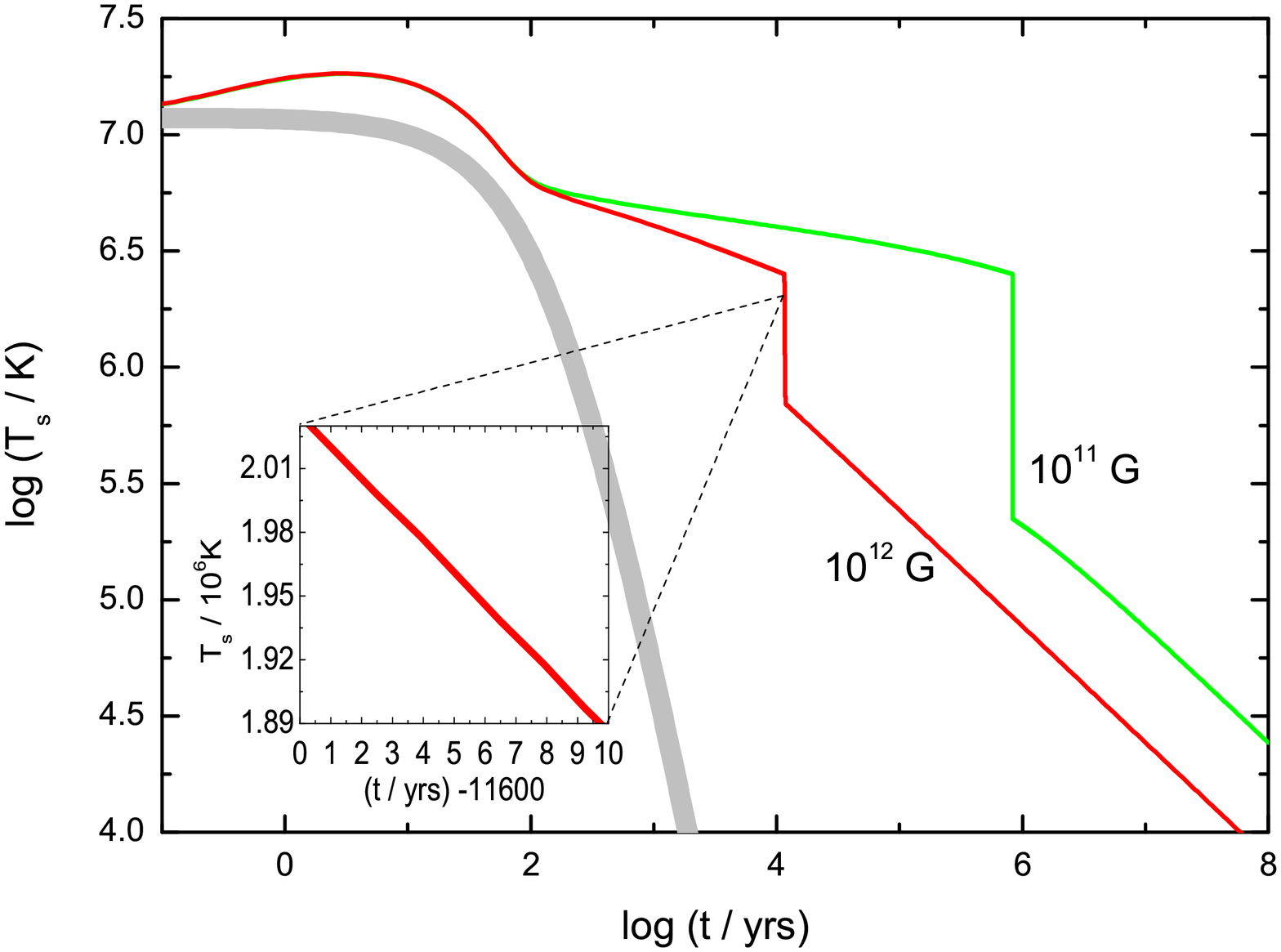}}
         \caption{The evolutions of the surface temperature of NSSs (left panel)
         and CSSs (right panel) with both DC and RM heating effects. The magnetic fields
         are the same as in Fig. \ref{fig:1}. The light grey
         bands present the case without heating.
         }
   \label{fig:3}
\end{figure*}

In Fig. \ref{fig:1}, we plot the spin-down behaviors of both NSSs
and CSSs. By considering that only a small fraction of quarks in the
2SC phase are paired, the evolution behavior of a 2SS is
qualitatively similar to a NSS (see \cite{Yu:2006}). So in this
paper the case of 2SSs will not be discussed specifically. As
previously found in \cite{Zheng:2006}, the spin velocity of SSs at
the very first time could decrease exponentially due to the
exponential increase of the r-modes. After the r-modes enter into
the saturation state with a constant amplitude, the spin evolution
dominated by the GR braking can be determined by
$\dot{\Omega}\propto -\Omega/\tau_g\propto-\Omega^{7}$, which gives
$\Omega\propto t^{-1/6}$. This analytical estimation could be pretty
valid for CSSs, which is similar to the case of neutron stars
\textbf{\cite{Owen:1998,Andersson:2001}}, but not so good for NSSs
for $t>100$ yrs due to a self-decrease of the saturation amplitude
(Instead, $\Omega\propto t^{-1/14}$ approximatively). After some
time, the braking due to MDR would become comparable to and
eventually exceed the GR braking, which leads to $\Omega\propto
t^{-1/2}$. In Fig. \ref{fig:1}, two values of the saturation of
$\kappa=10^{-6}$ and $10^{-8}$ are considered for a comparison.
Obviously, by pushing the saturation to smaller values, the
influence of the r-modes becomes weaker. Then, the spin evolution
returns to the traditional one due to only MDR and all issues
discussed in this paper return to the case considered in
\cite{Yu:2006}. So in the following calculations we only taken the
value $\kappa=10^{-6}$ into account. The other model parameters are
taken as follows. The initial values of the r-mode amplitude,
angular velocity, and interior temperature are $\alpha_0=10^{-10}$,
$\Omega_0=\frac{2}{3}\sqrt{\pi G \bar{\rho}}$, and $T_0=10^{10}$ K,
respectively. The parameters for SSs are $M=1.4M_{\odot}$, and
$R=10$ km, and for the SQM are $\alpha_c=0.2$, $\Delta=100$ MeV,
$m_s=100$ MeV, $q_n\sim20$ MeV, and $Y_e=10^{-5}$.

\begin{figure*}
\centering
\resizebox{\hsize}{!}{\includegraphics{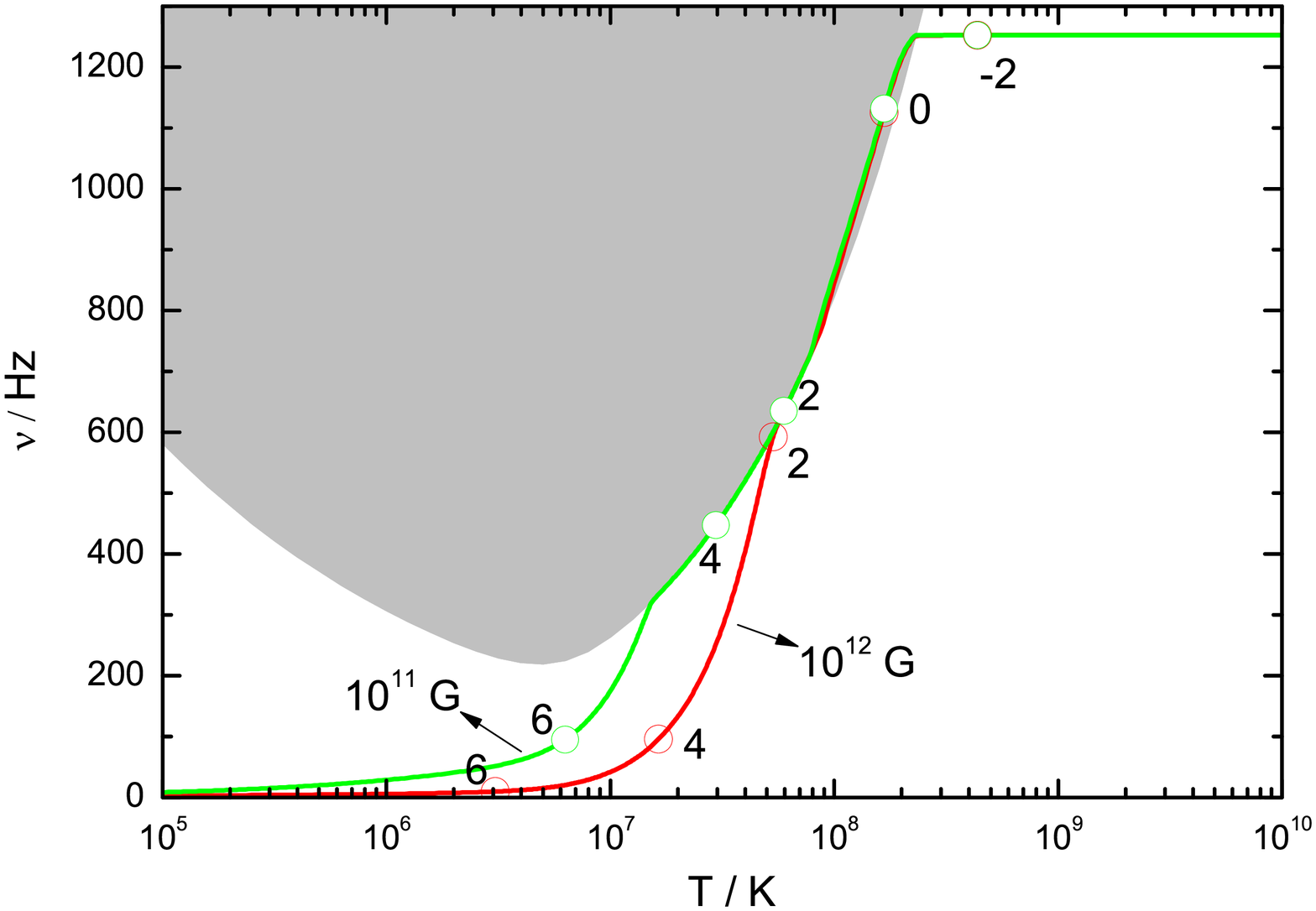}\includegraphics{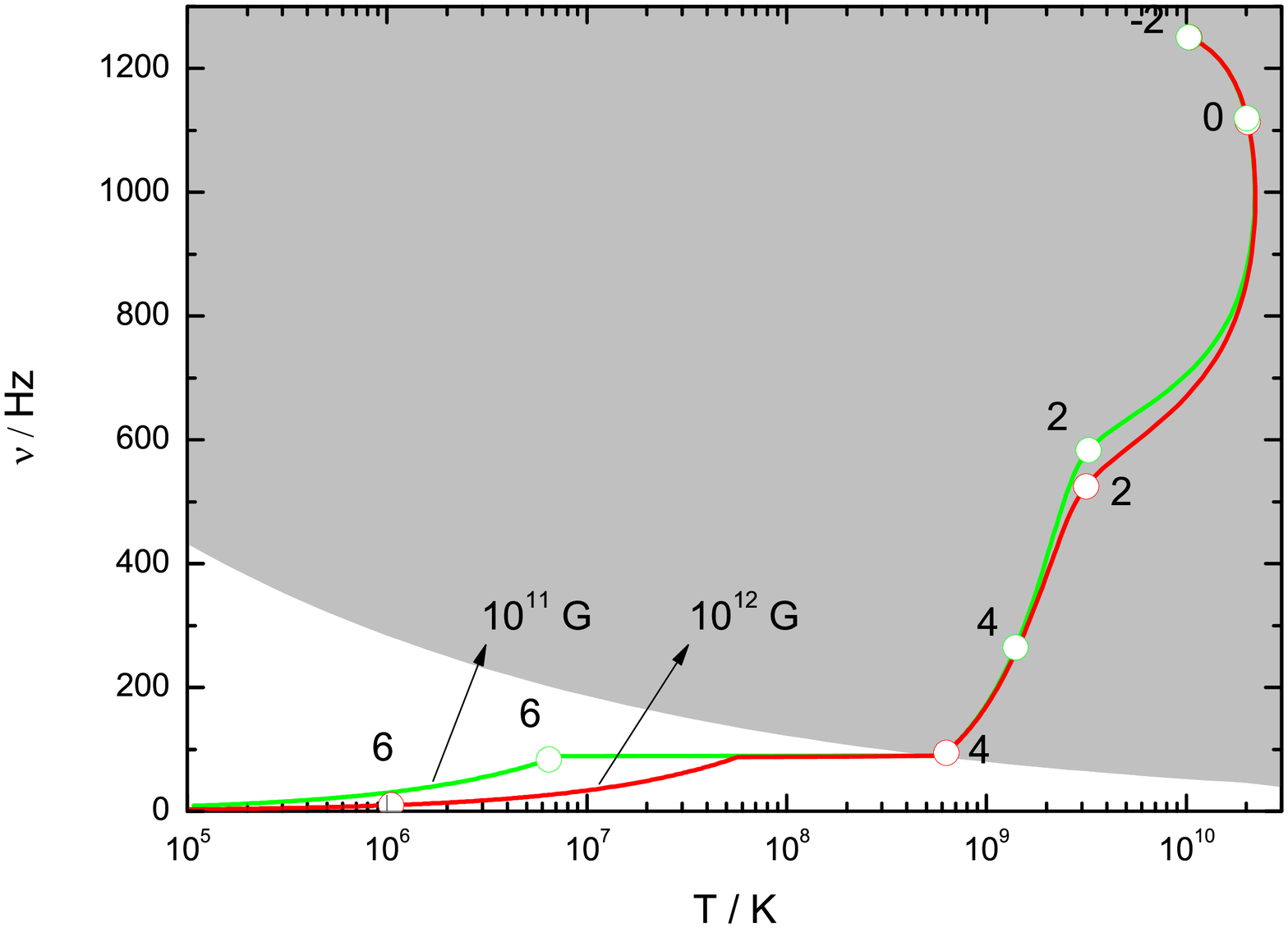}}
         \caption{The evolution of NSSs (left panel) and CSSs (right
         panel) for $\alpha_{\rm sat}=10^{-3}$. The shaded areas are r-mode
         unstable regions. The labels for the open cycles represent the value
         of log$(t/\rm yrs)$. The magnetic fields are the same as in Fig. \ref{fig:1}.}
   \label{fig:4}
\end{figure*}

\section{RESULTS} \label{Section IV}
The temporal dependences of the heating powers are numerically
calculated and presented in Fig. \ref{fig:2}. As shown, the RM
heating could paly a dominate role in the thermal evolution of SSs
during their middle ages, i.e., from $\sim100$ yrs to $\sim
10^{4-6}$ yrs. By contrast, the DC heating exceeds the RM heating
for $t\lesssim 100$ yrs or $t\gtrsim10^{4-6}$ yrs. The stronger the
dipole magnetic field, the weaker the RM heating. For NSSs with
$B=10^{12}$ G, the RM heating becomes always subordinate to the DC
heating. On one hand, following the analytical spin-down history of
CSSs, the temporal evolution of the DC heating can be approximated
to $H_{\rm DC}\propto \Omega \dot{\Omega}\propto t^{-4/3}$ in the GR
braking stage and $H_{\rm DC}\propto \Omega \dot{\Omega}\propto
t^{-2}$ in the MDR braking stage, while for the RM heating $H_{\rm
RM}\propto\Omega^2/\tau_{\rm sr}\propto t^{-5/12} T^{-1}$ and
$\propto t^{-5/4} T^{-1}$ in the GR and MDR braking stages,
respectively. On the other hand, for NSSs, we can get $H_{\rm
DC}\propto t^{-8/7}$ and following $H_{\rm DC}\propto t^{-2}$, while
$H_{\rm RM}\propto\Omega^2/\tau_{\rm sv}\propto t^{-1/7} T^{-5/3}$.
The RM heating is eventually switched off due to the end of the
r-mode instability. As a result, we plot the thermal evolution
curves of both NSSs and CSSs in Fig. \ref{fig:3}, where the standard
cooling of SSs without heating is shown by the light grey band. As
seen, the change in the thermal history of SSs arising from heating
effects could be remarkable, as previously claimed by
\cite{Yu:2006}.

For CSSs, as analyzed by Yu and Zheng \cite{Yu:2006}, the main
thermal evolution can be directly determined by the equilibrium
equation $L_{\gamma}=H$, since both $L_{\nu}$ and $C$ vanish due to
the color superconductivity. Therefore, their thermal history during
the equilibrium can be divided into RM and DC heating stages. In the
first stage, $T^{2}\propto t^{-5/12}T^{-1}$ (GR braking) and
$T^{2}\propto t^{-5/4}T^{-1}$ (MDR braking) gives $T\propto
t^{-5/36}$ and $T\propto t^{-5/12}$, respectively. The realistic one
could fall between these two extremes. In the second stage, the DC
heating and MDR braking gives $T\propto t^{-1/2}$. Here, the most
interesting thing could be the sharp transition between these two
stages. As clearly shown in the right panel of Fig. \ref{fig:3}, an
extremely steep decay of the stellar temperature appears in the
thermal evolution curves of CSSs at the age of $\sim10^{4-6}$ yrs.
The range of the temperature during this transition is
$\sim3\times10^5$ K$-3\times10^6$ K. Such a time and temperature
region in the $T-t$ panel is just consistent with the most
observational data \cite{Page:2004}. On the other hand, such a steep
decay indicates that the stellar temperature can be found to
decrease, most rapidly, by about seven percent in ten years, as
shown in the insert in the right panel of Fig. \ref{fig:3}. Another
interesting feature in the thermal evolution curves of CSSs is the
big bump appearing during the first hundred years, where the
equilibrium between the heating and cooling effects has not been
built ($H_{\rm DC}\gg L_{\gamma}$). Because of the appearance of the
r-modes, the deconfinement transition of the stellar crust can take
place much earlier and more quickly than the case with only MDR
braking, which means that most binding energy of the crust could be
released within the first hundred years. Therefore, the bumps
obtained here are nearly independent of the strength of the magnetic
fields, whereas in \cite{Yu:2006} such a bump only exists in the
high magnetic field case. It should be pointed out that, due to the
thermal relaxation of the stellar crust which is not considered in
our calculations, the variation of the interior temperature of SSs
at early ages actually cannot be immediately reflected by the
surface emission. However, because the crust of SSs is only tiny,
which is just the outer part of the crust of traditional neutron
stars, the relaxation time scale of SSs could only be on the order
of several years (e.g., see Figs. 2 and 3 in \cite{Schaab:1997b}).
Therefore, an infant CSS at the age of several tens to a hundred
years is still expected to have a high temperature of $\sim10^7$ K
(emitting at $\sim1$ keV).

For NSSs, their thermal histories are not very different from the
standard one, except for very late ages. In principle, the histories
of NSSs could be divided into four stages, which are determined by
$C\dot{T}=H-L_{\nu}$, $C\dot{T}=L_{\nu}$, $C\dot{T}=L_{\gamma}$, and
$C\dot{T}=H_{\rm DC}$ (MDR braking), respectively. Of course, for a
specific NSS with a fixed magnetic field, there may only be two or
three stages. In the first stage, the heating powers and the
neutrino luminosity could be comparable to each other, so an
analytical calculation is infeasible. For the following three
stages, we can easily get $T\propto t^{-1/4}$, $T\propto \exp(-t)$,
and $T\propto t^{-1/2}$, respectively, for $C\propto T$ and
$L_\nu\propto T^6$ (QDU). Different from CSSs, it is not easy to
produce an early temperature bump by NSSs because of their very
intense neutrino emission during the first hundred years.

Finally, for an integrated impression of the thermospin evolution of
SSs, we plot the evolution curves of SSs in the $T-\nu$ panel in
Fig. \ref{fig:4}, where the r-mode instability windows are also
presented. Some qualitative descriptions for such figures are
provided in \cite{Zheng:2006}. Here we emphasize that (i) for NSSs,
after their birth at the region with high $\nu$ and $T$ where the
r-mode is stable, their thermospin evolution is basically determined
by the boundary of the r-mode instability window and that (ii) for
CSSs, they can stay at the bottom of the r-mode instability window
for tens of thousands of years.

\section{SUMMARY AND DISCUSSIONS} \label{Section V}
The SS hypothesis and the specific state of the SQM in SSs are some
of the most fascinating mysteries in both physics and astronomy. In
this paper, we investigated the thermal evolution of SSs by
considering both the deconfinement transition of the stellar crust
and the r-mode instability of the fluid core. The results show the
following: (i) The cooling of NSSs with heating effects is only
somewhat different from the well-known standard one. Therefore,
there seems still no significant discrepancy between the NSS model
and the temperature observations of pulsars. (ii) For CSSs, there
are two interesting features found in their thermal histories. One
is the big temperature bump during the first hundred years and the
other is the steep decay of the temperature at the ages of
$\sim10^4$--$10^6$ yrs.

The serious problem of the CSS model is that, to our knowledge,
neither the early bump nor the steep decay ($\sim$7\% in ten years)
has been reported from observations, though they should be easily
detected if they indeed exist. Let us recall the constraint on the
CSS model given by Madsen \cite{Madsen:2000}, who claimed that the
CSS model is not permitted by the most rapid pulsars because these
pulsars locate deep at the r-mode instability window of CSSs.
Therefore, we conclude that both the rotational and thermal tests on
the SS models, which are independent of each other, suggest that the
CSS model is disfavored by both the rotation and temperature
observations of compact stars. As an alternative consideration
discussed in \cite{Madsen:2000}, the following two possibilities are
worthy of attention: (i) The CFL phase is not reached at densities
relevant in pulsars, but the SQM could still be stable because the
NSS model is not ruled out. (ii) The CFL phase appears in a hybrid
star, in which only the stellar core consists of quark matter.
Similar to the traditional neutron stars, hybrid stars could pass
the tests by both the spin and temperature observations of pulsars.

\acknowledgements We thank the anonymous referee for his comments,
which led to some essential discussions in this paper. This work is
supported by the National Natural Science Foundation of China
(Grants No. 11103004, No. 11073008, and No. 11178001), the
Foundation for the Authors of National Excellent Doctoral
Dissertations of China (Grant No. 201225), and the Self-Determined
Research Funds of CCNU from the college's basic research and
operation of MOE of China (Grant No. CCNU12A01010).

\end{document}